\documentclass{jpsj2}
\title{Dynamics of fluctuation of the top location of a sandpile}

\author{Chiyori \textsc{Urabe}\thanks{E-mail address: chiyori@yuragi.jinkan.kyoto-u.ac.jp}}

\inst{Graduate School of Human and Environmental, Kyoto University,
Kyoto 606-8501}

\abst{We investigate the fluctuation of the top location of a sandpile
 numerically using the two-dimensional discrete elements method. We feed
 particles to a sandpile at a fixed time interval and calculate power
 spectra from the time series of the top location. We find that the
 power spectra are approximated as power functions, and their exponents
 increase to $-1$ through a plateau as the time interval decreases. For
 small time interval, avalanches occur continuously either on the left
 or right side slope of a sandpile, and the slope on which avalanches
 take place switches intermittently. The long time fluctuation of the
 top location corresponds to the switchings. For the time series of the
 switchings, we discuss the relation between the power spectrum and the
 distribution of waiting times based on analytic theories.}

\kword{top location, fluctuation, sandpile, granular flow, avalanches, power law, switchings}

\begin{document}
\maketitle

\section{Introduction} %% No sections necessary for express letters, letters and short notes

Dense systems of granular materials exhibit solid-like and fluid-like
behaviors \cite{NeddermanB1992,JaegerNagelBehringerRMP68(1996),
JaegerNagelBehringerPT49(1996),DuranB2000}, and many researches are
devoted on them.
For the static phase, the spatial distribution of stress on the bottom of a
sandpile was measured in experiments, and it is known that its
functional form depends on the history of the formation of the sandpile 
\cite{WittmerClaudinCatesBouchaudN382(1996),VanelHowellClarkBehringerClementPRE60(1999),GengHowellLonghiBehringerPRL87(2001),GengLonghiBehringerHowellPRE64(2001)}.
For the fluid phase, steady flows and avalanches are investigated from
several perspectives
\cite{FretteChristensenMalthe-SorenssenFederJossangMeakinN379(1996),
NoijeErnstPRL79(1997),
BreyDuftyKimSantosPRE(1998),
DaerrDouadyN399(1999),
KadanoffRMP71(1999),
PouliquenPF(1999),
SilbertErtasGreatHalseyLevinePlimptonPRE64(2001),
MitaraiHayakawaNakanishiPRL88(2002),
GoldhirschARFM35(2003),
MitaraiNakanishiJFM507(2004)}.
Fluidized phase appears in a localized layer near the surface, and the
particle velocity under the layer obeys an exponential function of the
depth from the surface
\cite{KomatsuInagakiNakagawaPRL86(2001)}.
It is known for granular flows in a pipe that the temporal power spectra
of particle density obey power laws
\cite{PengHerrmannPRE49(1994),
PengHerrmannPRE51(1995),
HorikawaNakaharaNakayamaMatsushitaJPSJ64(1995),
HorikawaIsodaNakayamaNakaharaMatsushitaPhysA233(1996),
NakaharaIsodaPRE55(1997),
HayakawaNakanishiPTPS130(1998),
MoriyamaKuroiwaMatsushitaHayakawaPRE80(1998),
MoriyamaKuroiwaIsodaAraiTatedaYamazakiMatsushitaTG01(2001),
YamazakiTatedaAwazuAraiMoriyamaMatsushitaJPSJ71(2002),
MoriyamaKuroiwaTatedaAraiAwazuYamazakiMatsushitaPrgTPS(2003),
HayakawaCondMat0503171(2004),
AwazuM(2005)}.
 
In systems of a sandpile, feeding particles at small feed rate, the
surface of a sandpile is kept in solid states except when avalanches
occur intermittently, and continuous flows appear as the feed rate
increases.
However, even in the case that the feed rate is rather large, it is
infrequent that the whole surface of a sandpile is kept in the fluid
phase, and the states of the surface change temporally or spatially
\cite{AltshulerRamosMartinezBatista-LeyvaRiveraBasslerPRL(2003)}. 
The time evolution of a sandpile is caused by complicated interactions
between avalanches and the shape of the surface.

A sandpile typically becomes mountain shaped with a top as it grows
\cite{AlonsoHerrmannPRL76(1996),GrasselliHerrmannOronZapperiGM2(2000)}.
Because particles on the surface run down from the top to the foot of a
sandpile, the top plays the role of a singular point in an average flow
of particles. 
In the formation process of a sandpile, the top location moves with time
and determines the global flow on the surface.
In this paper, we focus on the fluctuation of the top location to
characterize the long time evolution of a sandpile.
We carry out two-dimensional numerical simulations and measure the
power spectrum of the top location.
We find that the power spectrum obeys a power function, and that the
exponent depends on the feeding rate.

The organization of the remainder of this paper is as follows.
In \textsection \ref{MethodDEM}, we explain the method of our numerical
simulations. 
In \textsection \ref{results}, we investigate the power spectrum of the top
location and relations between its exponent and the feed rate of
particles.
In \textsection \ref{discussion}, we discuss the origin of this power law.
Finally, we draw conclusions in \textsection \ref{Conclusions}. 

\section{Method Of Simulations}
\label{MethodDEM}

\begin{figure}[b]
\begin{center}
\includegraphics[width=8cm]{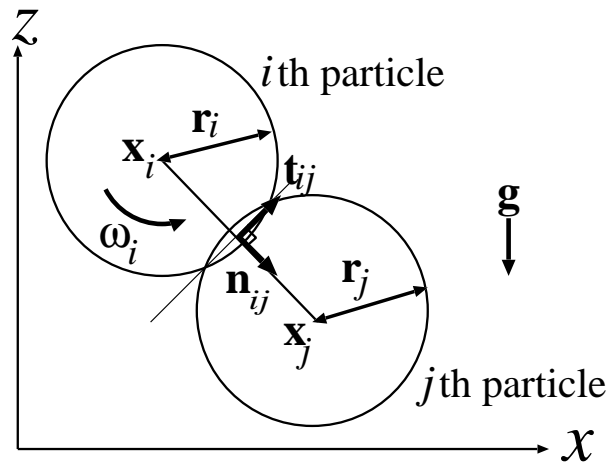}
\caption{\label{particles}  The $i$th particle and the $j$th particle
 are  in contact.}
\end{center}
\end{figure}

In order to simulate polydispersed circular particles, we adopt a
two-dimensional Discrete Elements Method
(DEM) \cite{CundallStrackG29(1979)}. 
In DEM, we assume that particles are cohesionless, and the mass density
of a particle per unit area, $\rho_{0}$, is constant.
The linear spring model is adopted to describe the repulsion of two
particles in contact, and the viscous forces and Coulomb slip are
assumed to act on them.
We assume that the $i$th particle has the radius $r_{i}$, the mass
$m_{i}=\pi r_{i}^{2}\rho_{0}$  and the momentum of inertia
$I_{i}=m_{i}r_{i}^{2}/2$.
The center of mass and the angular velocity of the $i$th  particle
represent $\mathbf{x}_{i}$ and $\omega_{i}$, respectively.
The equations of motion of the $i$th particle are given by
\begin{eqnarray} 
\begin{array}{rcl}
 m_{i}\ddot{\mathbf{x}}_{i}&=&\sum_{j}\Bigl(F_{n}^{ij} \mathbf{n}_{ij}+F_{t}^{ij}\mathbf{t}_{ij} \Bigr) + m_{i}\mathbf{g},\\
 I_{i}\dot{\omega}_{i}&=&r_{i}\sum_{j}F_{t}^{ij}, 
\end{array}
\end{eqnarray}
where $\mathbf{n}_{ij}$ and $\mathbf{t}_{ij}$ represent the normal and
the tangential unit vectors at the contact point between the $j$th particle
and the $i$th particle, and $\mathbf{g}$ is the acceleration of gravity, 
as depicted in Fig.\ \ref{particles}. 
The normal contact force $F_{n}^{ij}$ is defined by
\begin{equation}
F_{n}^{ij}=\tilde{F}_{n}^{ij}\Theta(-\tilde{F}_{n}^{ij}),
\end{equation}
and
\begin{equation}
\tilde{F}_{n}^{ij}=-k_{n}m_{r}^{ij}\Bigl(r_{i}+r_{j}-|\mathbf{x}_{i}-\mathbf{x}_{j}|\Bigr)-\eta m_{r}^{ij}\mathbf{n}_{ij}\cdot \Bigl(\dot{\mathbf{x}}_{i}-\dot{\mathbf{x}}_{j}\Bigr), 
\end{equation}
where $\Theta(x)$ is the Heaviside function, and $m_{r}^{ij}$ is the
reduced mass $m_{r}^{ij}\equiv\frac{m_{i}m_{j}}{m_{i}+m_{j}}$.
$\Theta(-x)$ is introduced so that the contact force is repulsive.
The tangential contact force $F_{t}^{ij}$ is defined by
\begin{equation}
F_{t}^{ij}=k_{t}m_{r}^{ij}u_{t}^{ij}.
\end{equation}
$u_{t}^{ij}$ is obtained by integrating the equation
\begin{equation}
\dot{u}_{t}^{ij}=-\Bigl( (\dot{\mathbf{x}}_{i}-\dot{\mathbf{x}}_{j})\cdot \mathbf{t}_{ij}+r_{i}\omega_{i}+r_{j}\omega_{j}\Bigr)\Theta\bigl(\mu|F_{n}^{ij}|-|F_{t}^{ij}| \bigr),
\end{equation}
where $u_{t}^{ij}$ is zero when the particles are not in contact (i. e. for
$|\mathbf{x}_{j}-\mathbf{x}_{i}|> r_{j}+r_{i}$).
Here, Coulomb frictional coefficient $\mu$ is assumed to be $0.5$.
We express the maximum diameter and  the maximum weight of a particle as
$d$ and $m$.
The distribution function of the diameters is uniform in the range
between $0.8d$ and $d$. 
We assume that the spring constants are $k_{n}=1.0\times 10^{4}mg/d$ in
the normal direction, and $k_{t}=2.0\times 10^{3}mg/d$ in the tangential
direction,  and that the viscosity is $\eta=1.0\times10^{2}\sqrt{g/d}$.    
The coefficient of restitution in our model is about $0.2$ for a head-on
particles collision.
We adopt the second-order Adams-Bashforth method for time-integration
with the time interval $\Delta t=1.0\times 10^{-3}\sqrt{d/g}$. 

\begin{figure}[t]
\begin{center}
\includegraphics[width=8cm]{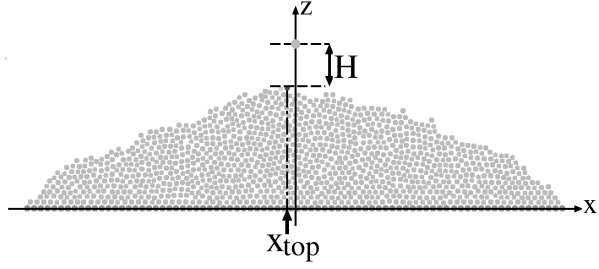}
\caption{\label{sandpile}Setup of a sandpile in our
 simulations. $x_{top}$ represents the horizontal top location. Fed
 particles are released at the height $H$ from the top.}
\end{center}
\end{figure}

We assume that a particle is in a sandpile if the particle contacts other
particles, and the top location of the sandpile is defined as the center
of mass of the highest particle in the sandpile (Fig.\ \ref{sandpile}).
The floor under a sandpile is a horizontal array of $80$ fixed particles
with diameter $d$.
We introduce the $x$ coordinate along the floor and define the origin at
the center of the floor. 

To investigate the fluctuation of the top in the formation process of a
sandpile, we drop particles with the time interval $T$ to it.
The particles are released at a position just above the center of the
floor and its height is $H$ from the top location as shown in
Fig.\ \ref{sandpile}.
We first make a sandpile grow until it covers the floor and use this 
sandpile as an initial state.
Because particles run off the edges of the floor with finite length, the
size of a sandpile is maintained almost constant.

After the time series of the top location $x_{top}(t)$ reaches a
statistical stationary state, we calculate its power spectrum $S(f)$
with respect to frequency $f$.
To calculate the power spectrum $S(f)$ from the time series, we divide
it into $M$ time series with a time interval $T^{(s)}$, the $m$th power
spectrum $S_{m}(f_{j})$ is defined by
\begin{eqnarray}
 S_{m}(f_{j})\equiv \frac{1}{N}|\sum_{n=1}^{N}x_{top,n}^{(m)}e^{-\frac{2\pi in}{N}j}|^{2}
 \label{Sm}
\end{eqnarray}
where $x_{top,n}^{(m)}\equiv x_{top}((m+\frac{n}{N})T^{(s)})$ and
$f_{j}\equiv \frac{j}{T^{(s)}}$.
 
We introduce the power spectrum $S(f)$ as the average of the $M$
power spectra,
\begin{eqnarray}
S(f_{j})\equiv \frac{1}{M}\sum_{m=0}^{M-1}S_{m}(f_{j}).
\label{S}
\end{eqnarray}

\section{Results}
\label{results}

\begin{figure}[t]
\begin{center}
\includegraphics[width=8cm]{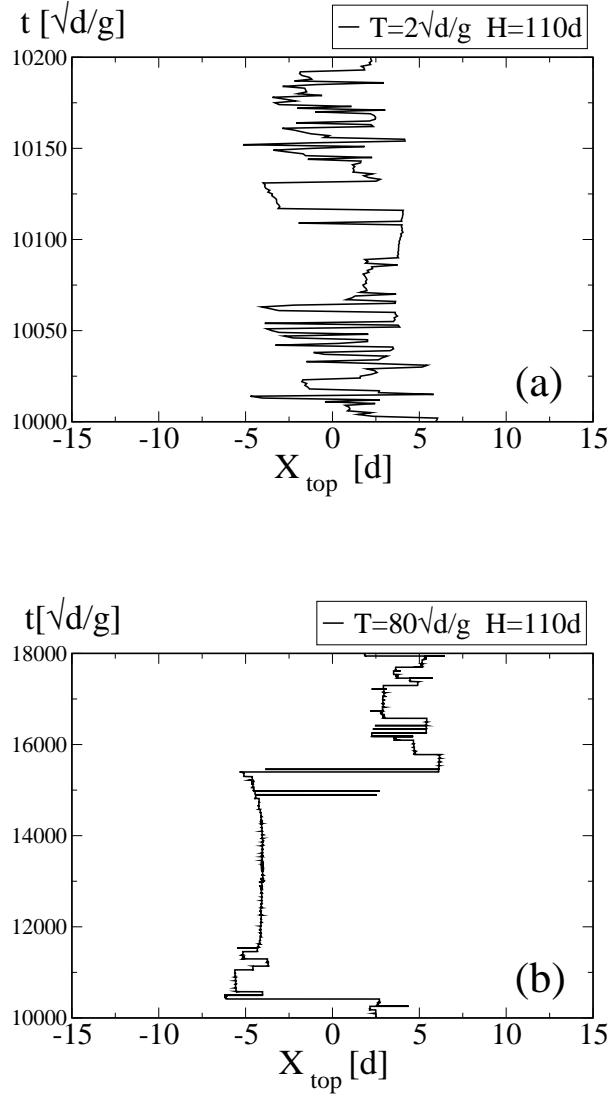}
\caption{\label{xH110} Time series of the top location $x_{top}(t)$. We
 use $H=110d$, $T=2\sqrt{d/g}$ (a) and $T=80\sqrt{d/g}$ (b). The
 vertical axes are time $t$. We note that the same number of particles
 are fed in (a) and (b).}
\end{center}
\end{figure}

\begin{figure}[t]
\begin{center}
\vspace*{1cm}
\includegraphics[width=8cm]{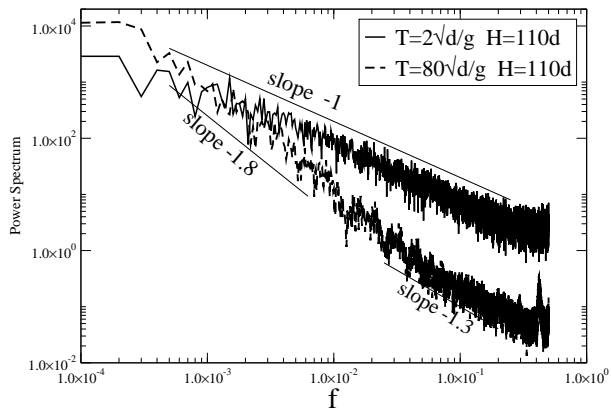}
\caption{\label{PSH110T02_80} Double logarithmic plot of $S(f)$ for the
 time series. $H=110d$, $T=2\sqrt{d/g}$ (black line) and $T=80\sqrt{d/g}$
 (dashed line). The lines are drawn for reference.}
\end{center}
\end{figure}

We measure $x_{top}$ for various  values of the time interval $T$ and
the height $H$.
We change $H$ in the range $20d \leq H \leq 110d$.
If $H$ is sufficiently large beyond this range, the impact of a dropped
particle is large and collapses the top shape of a sandpile into a
caldera \cite{GrasselliHerrmannGM3(2001)}.
Figures\ \ref{xH110}(a) and \ref{xH110}(b) shows the time series of $x_{top}(t)$
obtained from simulations with $T=2\sqrt{d/g}$ (a) and $T=80\sqrt{d/g}$
(b).
The top fluctuates frequently in the case of small $T$, on the other
hand, in the case of large $T$, the top almost stays for long time in
comparison with $T$ because the motion of particles induced by the
impact of a fed particle ceases before the next particle is dropped.
Figure\ \ref{PSH110T02_80} is the power spectra of the time series,
$S(f)$, which is calculated using eqs. (\ref{Sm}) and (\ref{S}) with
$T^{(s)}=N=10000$ and $M=10$.
We find that $S(f)$ behaves as a power-law, and its exponent depends on
the value of $T$.
At $T=2\sqrt{d/g}$, $S(f)$ approximately obeys $1/f$ law.
At $T=80\sqrt{d/g}$, the exponents of $S(f)$ are smaller than $-1$.

\begin{figure}[t]
\begin{center}
\vspace*{1cm}
\includegraphics[width=8cm]{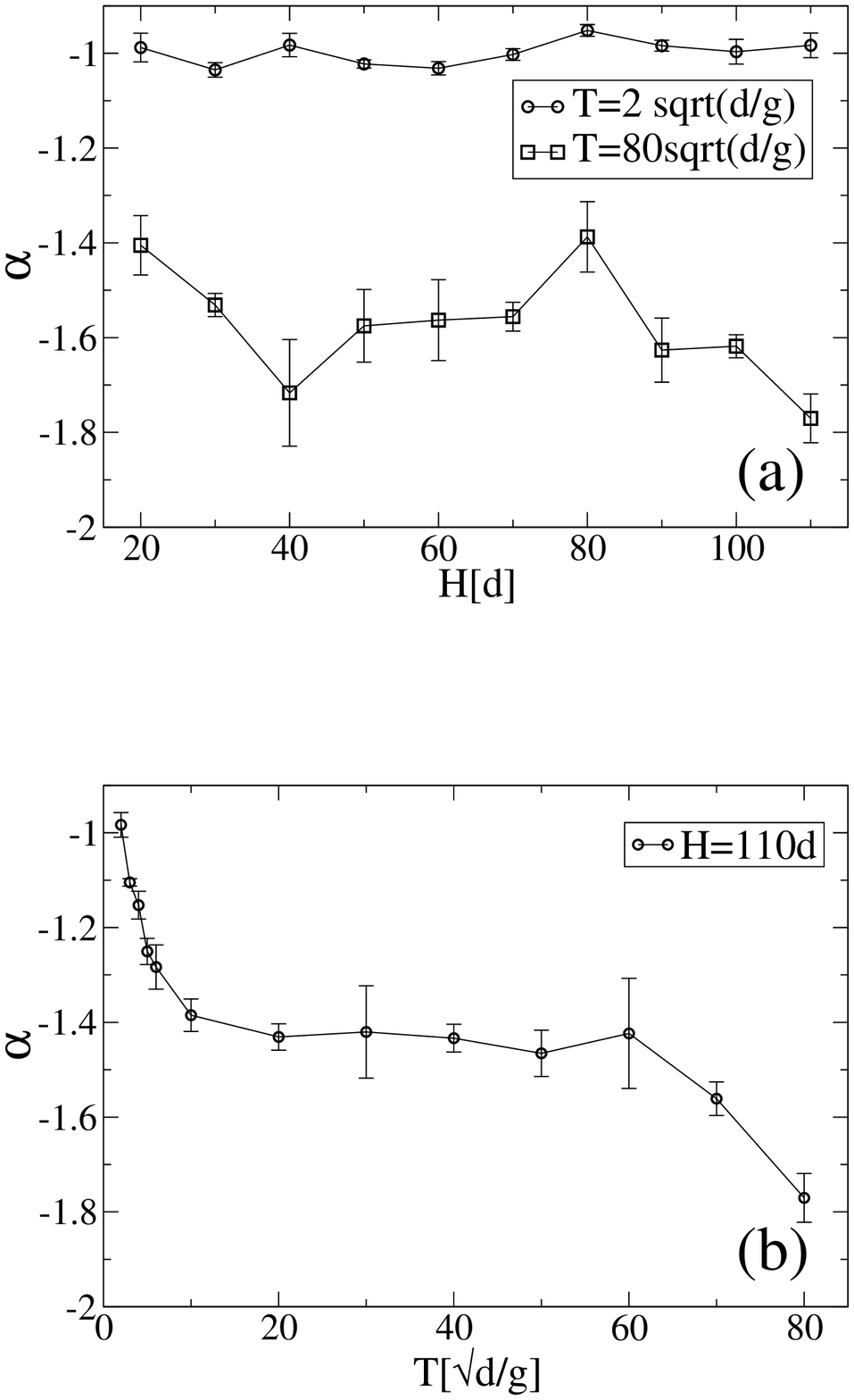}
\caption{\label{alpha} (a) Dependence of $\alpha$ on $H$ and (b)
 dependence of $\alpha$ on $T$. The error bars correspond to the
 standard deviations.}  
\end{center}
\end{figure}

We investigate the dependence of the exponent of $S(f)$, $\alpha$, on
$T$ and $H$.
For the range $5/T^{(s)}\leq f \leq 1/(2T)$, we calculate $\alpha$ from
the double logarithmic plot of $S(f)$ in the least square method.
$\alpha$ is insensitive to $H$ as shown in  Fig.\ \ref{alpha}(a).
In contrast, Fig.\ \ref{alpha}(b) indicates that $\alpha$ strongly
depends on $T$ for small $T$ and approaches $-1$ as $T$ decreases.
As $\alpha$ approaches $-1$, the power spectrum is approximated as a
power function with a high degree of accuracy as indicated with the
error bars.
In the range $10\sqrt{d/g}<T<60\sqrt{d/g}$, $\alpha$ is approximately a
 constant $-1.43\pm0.03$, although $\alpha$ decreases as  $T$ increases
 beyond $60\sqrt{d/g}$.  
In the case of large $T$, the error bars in Figs.\ \ref{alpha} are large
because the range of frequency used for fitting is small. 
In the region of higher frequency than $1/T$, the power function with
the same exponent is not best fit with $S(f)$.
If we fit $S(f)$ with a power function in this range of high frequency,
its exponent changes from that indicated in Figs.\ \ref{alpha}, as shown
with the data of $T=80\sqrt{d/g}$ in Fig.\ \ref{PSH110T02_80}. 

\begin{figure}[t]
\begin{center}
\vspace*{1cm}
\includegraphics[width=8cm]{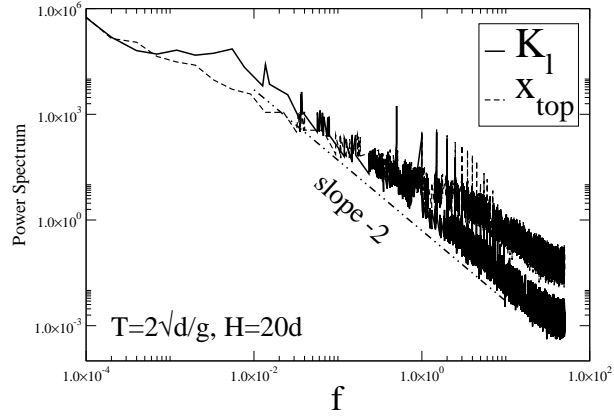}
\caption{\label{PSxkL} Power spectra of the time series of $x_{top}(t)$
 and $K_{l}(t)$ for $T=2\sqrt{d/g}$ and $H=20d$. The line with a slope is
 drawn for reference.}
\end{center}
\end{figure}
 
\begin{figure}[t]
\begin{center}
\includegraphics[width=8cm]{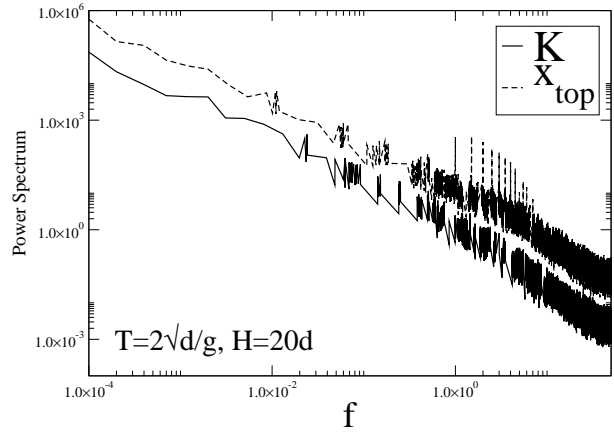}
\caption{\label{PSxK} Power spectra of the time series of $x_{top}(t)$
 and $K(t)$ for $T=2\sqrt{d/g}$ and $H=20d$.}  
\end{center}
\end{figure}

We mainly focus on the case of small $T$ because we are, in particular,
interested in the case that $\alpha$ is close to $-1$.
The displacement of the top location $x_{top}$ is caused by avalanches,
and avalanches occur on either slope at almost all times in this case.
Although the instantaneous magnitude of avalanches is characterized by the
kinetic energy of the particles, $S(f)$ and the power spectrum of the
kinetic energy differ in functional form as mentioned below.
Eliminating the narrow region $-d \leq x \leq d$ in the center of a
sandpile, we divide the sandpile into the left part and the right part
with respect to $x=0$.
We measure the kinetic energies of the left and right parts, $K_{l}(t)$
and $K_{r}(t)$, respectively.
Using the same definition in eqs. (\ref{Sm}) and (\ref{S}), the power
spectra of the time series of $K_{l}(t)$ and $K_{r}(t)$ are calculated
with $T^{(s)}=1.0\times 10^{4}\sqrt{d/g}$, $N=1.0\times 10^{5}$ and
$M=10$.
For $T=2\sqrt{d/g}$ and $H=20d$, the power spectra of both
$K_{l}(t)$ and $x_{top}(t)$ are shown in Fig.\ \ref{PSxkL}.
The power spectrum of $K_{r}(t)$ is  similar to that of $K_{l}(t)$.
Because the power spectra of $K_{l}(t)$ and $K_{r}(t)$ are
Lorentzian-like, avalanches seem to occur at random. 

From the results of numerical simulations, it is found to be rare that
avalanches occur simultaneously on the left and right slopes of a
sandpile.
We refer the states that avalanches occur on the left and right slopes
as left mode and right mode, respectively.   
To investigate switchings between the both mode, we define the binarized
time series,
\begin{equation}
K(t)=
\begin{cases}
      +1, & \text{for $K_{l}(t)<K_{r}(t)$}, \\
      -1, & \text{for $K_{l}(t)\geq K_{r}(t)$}.  
\end{cases}\label{K}
\end{equation}
The sign of $K(t)$ represents the side on which avalanches occur mainly
at time $t$.
The switchings are well defined by $K(t)$ in the case that $T$ is small.
However, as $T$ increases, it is difficult to define the
switchings because avalanches occur at intervals, and the time intervals
between avalanches are comparable to the time scale of switchings. 
We find that the time series of $K(t)$ is similar to $x_{top}(t)$ for
small $T$.
The power spectrum of the time series $K(t)$ is shown in Fig.\ \ref{PSxK}
for $T=2\sqrt{d/g}$ and $H=20d$.
The power spectrum of $K(t)$ is approximated as a power law with the
exponent of $-1$ in the long time scale.
The exponent is approximately the same as that of the top location.
Investigating the conditional probability of $K(t)=-1$ for a given
$x_{top}$, this probability increases with the value of $x_{top}$.
Therefore, in each mode, the top location $x_{top}(t)$ is mainly in the
opposite side on which avalanches occur.
Thus the fluctuation of $x_{top}$ corresponds to the switching between the
two modes, but not to the fluctuation of the magnitude of avalanches.

\section{Discussion}
\label{discussion}

\begin{figure}[t]
\begin{center}
\vspace*{1cm}
\includegraphics[width=8cm]{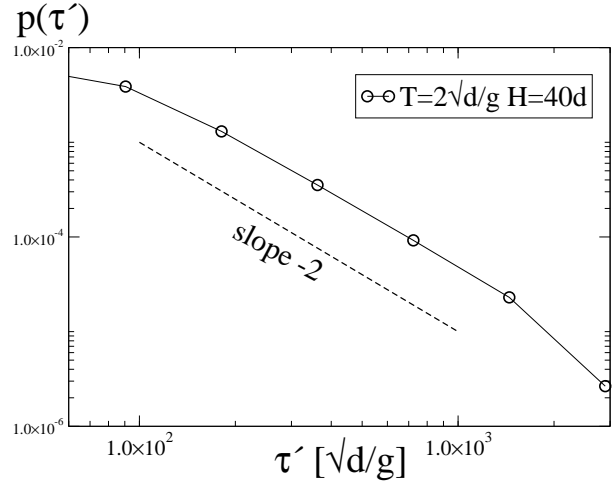}
\caption{\label{P} The distribution of the waiting time in $K(t)$ for
 $T=2\sqrt{d/g}$ and $H=40d$.} 
\end{center}
\end{figure}

For the binarized time series such as $K(t)$ defined by eq. (\ref{K}), it
is known from an analytical theory that its spectrum is expressed as a
power function if the waiting time has a power-law distribution and each
interval is independent
\cite{LowenTeichPRE47(1993),FuchikamiIshiokaProcSPIE5471(2004)}.
Here, the waiting time $\tau$ is defined as a time interval between
neighboring switchings in the binarized time series.
We assume that the probability density of $\tau$, $p(\tau)$, is the
abrupt-cutoff power law,  
\begin{eqnarray}
p(\tau)= \begin{cases}
c t^{D}, & \text{for $a<t<b$},\\
0, & \text{otherwise},
\end{cases}
\end{eqnarray}
where the constants $a$ and $b$ are sufficiently small and large
respectively, and $c$ is the normalization constant.
In the range of $1/b\ll f\ll 1/a$, the power spectrum of this binarized
time series, $S_{b}(f)$, is given approximately by
\begin{eqnarray}
S_{b}(f)\sim f^{-(D+3)}
\end{eqnarray} 
for $-3<D<-1$.
In the case of small $T$, $p(\tau)$ is expected to be a power function
with $D\cong -2$ because the exponent of the power spectrum of $K(t)$ is
approximately $-1$ in our simulations.
However there are few intervals with waiting time longer than $\tau\cong
100\sqrt{d/g}$ in $K(t)$ because small noises chop up long intervals.
Therefore applying the median filter of a time width of $60\sqrt{d/g}$
to $K(t)$, we calculate the distribution of waiting time for the
coarse-grained time series, $p(\tau^{'})$. 
We find that $p(\tau^{'})$ decays approximately as a power function
with $D\cong -2$ as shown in Fig.\ \ref{P}.

The plateau with $\alpha\cong -1.4$ appears in a wide range of $T$ as
shown in Fig.\ \ref{alpha}(a), although the switchings can not be
well-defined by $K(t)$ as $T$ increases.
It is possible that the dynamics of the top location has
some relations with the density fluctuation of flows. 
In experiments of granular flow in vertical pipes filled with fluid, the
exponents  $-1$, $-4/3$ and $-3/2$ are reported for the temporal power
spectra of density 
\cite{
HorikawaNakaharaNakayamaMatsushitaJPSJ64(1995),
HorikawaIsodaNakayamaNakaharaMatsushitaPhysA233(1996),
NakaharaIsodaPRE55(1997),
MoriyamaKuroiwaMatsushitaHayakawaPRE80(1998),
MoriyamaKuroiwaIsodaAraiTatedaYamazakiMatsushitaTG01(2001),
YamazakiTatedaAwazuAraiMoriyamaMatsushitaJPSJ71(2002),
MoriyamaKuroiwaTatedaAraiAwazuYamazakiMatsushitaPrgTPS(2003),
AwazuM(2005)}, and the exponents $-1.4$ and $-3/2$  appear in traffic
flows 
\cite{MushaHiguchiJJAP15(1976),MushaHiguchiJJPA17(1978),
TakesueMitsudoHayakawaPRE68(2003)}.
We note that these exponents are close to $-1.4$.

As $T$ becomes sufficiently large, we believe that $\alpha$ approaches $-2$.
The top location stays at the same place for long time in comparison with
$T$, and its displacements are caused by impulsive force of avalanches.  
We infer that the top location moves like as a random walk, and its
power spectrum is Lorentzian-like, which decays as $f^{-2}$.

\section{Conclusions}
\label{Conclusions}

Carrying out 2-D DEM simulations, we have investigated the fluctuation
of the top location of a sandpile that is caused by avalanches and
piling up particles.
We have found that the power spectra of the time series of the top location
$x_{top}(t)$ behave as power functions in the range of long time scale.
The exponent of the power spectrum, $\alpha$, depends on the time interval
$T$ at which particles are fed to the sandpile.
$\alpha$ is close to $-1$ for small $T$ and decreases through a plateau
with $\alpha\cong -1.4$ as $T$ increases.
In the case of $\alpha\cong -1$, avalanches occur mainly either on the
left or right side slopes, and the states of the sandpile switch
intermittently between the left and right modes.
The power spectrum of the top location is approximately the same as that
of the binarized time series defined from the switchings.
In our simulations, the distribution of waiting time of the switchings
obeys a power function with the exponent $D\cong -2$ in this case.
The relation between $D$ and $\alpha$ is consistent with the
equation $\alpha+D=-3$ proposed in the analytical theories
\cite{LowenTeichPRE47(1993),FuchikamiIshiokaProcSPIE5471(2004)}.

\section*{Acknowledgment}

C. Urabe thanks H. Hayakawa, H. Tomita, S. Takesue and S. Kitsunezaki
for fruitful discussion and N. Fuchikami for discussion on the binarized
 time series \cite{FuchikamiIshiokaProcSPIE5471(2004)}. The author thanks
 H. Nakao for providing information on the paper of S. B. Lowen and
 M. C. Teich \cite{LowenTeichPRE47(1993)}. The author appreciates
 H. Hayakawa and S. Kitsunezaki for their critical reading. This work is
 partially supported by Grant-in-Aids for Japan Space Forum and
 Scientific Research (Grant No. 15540393) of the Ministry of Education,
 Culture, Sports, Science and Technology, Japan.

\end{document}